# Two-photon emission from thin superconducting rings irradiated by coherent microwave fields

A.I. Agafonov
NIC "Kurchatov Institute"
123182 Moscow, Russia

*Abstract*—Superconducting current instability in thin rings, irradiated by a coherent **microwave electromagnetic field**, is studied. On the regular circular motion of the Cooper pairs in the **supercurrent states the field** imposes coherent oscillations of the condensate as a whole. The oscillating condensate **should** emit photons with the energies determined **mainly** by the discrete values of the superconducting ring energy. For the first time the probability of the supercurrent decay accompanied by the two-photon emission, is derived. Numerical results for the energy-angular distributions of the photons, as well as **the waiting time of the stimulated two-photon transition** depending on both the ring sizes and **fluxoid** number in the initial current state, are presented.

***Keywords-component: two-photon emission, superconducting ring***

## I. INTRODUCTION

The microwave field-induced decay of the supercurrent in thin rings with single-photon emission has been predicted in [1]. This effect can be observed only in the thin-film rings, the thickness of which is less than both the field skin-depth and the London penetration depth. Under this condition the coherent field can cause the collective transition of all the Cooper pairs involved in the supercurrent. The latter decays by quantum jumps that correspond to destruction of one or several the magnetic-flux quanta (fluxoids) trapped in the ring. Of course, in massive rings the effect cannot occur and the supercurrent is persistent.

The first-order perturbation term of the electromagnetic interaction operator

$$\hat{V} = \int \hat{\mathbf{j}} \hat{\mathbf{A}} d\mathbf{r}, \quad (1)$$

where $\hat{\mathbf{j}}$ is the operator of the superconducting current density in the ring, $\hat{\mathbf{A}}$ is the vector-potential of the electromagnetic field generated by the ring, and the integration is taken over the ring volume $\Omega_r$, is inversely proportional to $\Omega_r$, as was shown in [1]. Hence the composite matrix element of (1) which describes the sequential emission of the two photons by the ring, $\propto \Omega_r^{-2}$. The matrix element of the electromagnetic interaction, which is proportional to $\hat{\mathbf{A}}^2$ and describes the simultaneous emission of two photons, $\propto \Omega_r^{-1}$, as we see below. Thus, for macroscopic rings the two-photon emission will be determined by the last interaction.

In the report we, based on the above motivation, study the microwave field-stimulated two-photon emission from the superconducting rings. We show that for the micron size rings this emission is characterized by larger probabilities as compared with single-photon one.

## II. THE CONDENSATE WAVE FUNCTION

Consider a thin superconducting ring with the rectangular cross section. Its inner radius $a \gg \lambda$ (where $\lambda$ is penetration depth) and the outer radius $b$ such that $b - a \gg \lambda$, and $(b-a)/b \ll 1$. The ring thickness $d$ is less than both the field skin-depth and the London penetration depth.

Properties of superconducting rings are completely determined by the condensate wave function $\Psi_m$, where $m$ is the number of the magnetic induction flux quanta (fluxoids), trapped in the ring. So, this function is defined the quantization of the magnetic flux $\Phi_m = m\Phi_0$, the discrete values of the total current $\Phi_m/L$, and the ring energy which is composed of the magnetic energy and the kinetic energy of the Cooper pairs, $E_m = E_0 m^2$, where $\Phi_0 = \pi\hbar/e$ is the fluxoid, $L$ is the ring self-inductance and $E_0 = \Phi_0^2/2L$ is the ring characteristic energy.

In a superconductor the identical wave functions for all the Cooper pairs are phase coherent. It leads to a macroscopic coherent state in which all the pairs are in the same quantum state with the well-defined phase. The wave function of a pair includes the wave function of the internal motion, wave function for the center of mass and the spin function of the pair. The first function has the characteristic length scale defined by the coherence length that $> 10^3$ angstrom for the BCS superconductors and $\leq 30$ angstrom for the HTSC materials. In the supercurrent state the second function is described by the phase function $\exp(i\phi_m)$.

The condensate wave function which contains the identical wave functions for all the Cooper pairs is asymmetric. For the singlet pairing it is achieved by its asymmetric spin part. The interaction of the condensate with electromagnetic fields does not affect on the spin variables. Therefore, without any restriction, the spin wave function of the condensate can be omitted. Further, we assume that the normalized wave functions of the internal motion of the pairs do not change at the supercurrent transitions. As a result, this phase function $\exp(i\phi_m)$ for each boson is important.

The superconducting condensate can be described as the superposition of states $|N>$ with different numbers of the Cooper pairs $N$. This superposition has the form of the coherent wave packet [2,3]:

$$\Psi_m = \sum_N c_N \psi_{Nm}(\mathbf{\rho})|N>, \qquad (2)$$

where $c_N$ is the probability amplitude of the state with $N$ particles, and the normalized condensate wave function for this state can be written as:

$$\psi_{Nm}(\mathbf{\rho}) = \frac{1}{\sqrt{\Omega_r}} \exp(iN\phi_m). \qquad (3)$$

Here $\Omega_r$ is the ring volume, $\phi_m = m\varphi$ is the phase of the single boson wave function, $\varphi$ is the azimuthal angle in cylindrical coordinates tied to the ring ( $a \le \rho \le b$ and $-d/2 \le z \le d/2$ ).

The amplitude $c_N$ is peaked up around the average value $\overline{N}$ in the ring. The width of this peak $\Delta N$ is the standard deviation of the number of bosons in the condensate, or, in other words, the uncertainty in their numbers. Of course, $\Delta N / \overline{N} \ll 1$. $N$ and $\phi_m$ are canonically conjugate variables. In general, $\Delta N$ is large and the phase may be treated as quasi classical [2]:

$$\frac{d}{dt}\hbar\nabla\phi_m = 2e\mathbf{A}_0 \cos(\omega_0 t)), \qquad (4)$$

where the right-hand side of (4) is the force induced by the coherent electromagnetic field irradiated the ring, $\mathbf{A}_0$ and $\omega_0$ are its amplitude and frequency $\omega_0$. The latter is assumed to be small as compared with the superconducting gap. Also, we imply that the wavelength of the field is much larger than the characteristic size of the ring, so that the field can be considered as uniform. Without loss of generality, the vector $\mathbf{A}_0$ is considered to be directed against the $y-$axis for the unit vector of which we have $\mathbf{i}_\varphi \mathbf{i}_y = \cos\varphi$.

From (4) with the initial condition $\phi_m(t=0) = m\varphi$, we find that the phase of the single boson wave function in the initial state of the ring is:

$$\phi_m(t) = m\varphi - \frac{2eA_0}{\hbar\omega_0}\rho\sin(\varphi)\sin(\omega_0 t). \qquad (5)$$

Of course, this field can change the ring energy in the initial state. To avoid this, we consider the field of low intensities, for which the field correction to the boson velocity $\mathbf{v} = \hbar\nabla\phi_m / m_C$ ($m_C$ is the mass of the Cooper pair) is small,

$$(\xi_0 b/m)^2 \ll 1, \qquad (6)$$

where

$$\xi_0 = \frac{2eA_0}{\hbar\omega_0}. \qquad (7)$$

Thus, in the case (6) the potential energy of the condensate in the external field $U_f = 2eN\mathbf{E}_0\mathbf{r}\cos(\omega_0 t)$ does not influence the ring energy, but only change the phase of the wave functions for the bosons (5).

### III. THE TRANSITION OPERATOR

We study the stimulated transition of the superconducting condensate with $N$ bosons from the initial state corresponding to the $m$ fluxoids, trapped in the ring, to the final state with $m_1(<m)$ fluxoids. This process accompanied by the two-photon emission, means the simultaneous transition of all the Cooper pairs. The Hamiltonian of the system can be expressed as:

$$H = H_0 + W + U_f, \qquad (8)$$

where $H_0$ is the Hamiltonian for the initial supercurrent state and free electromagnetic field,

$$W = \frac{2e^2 N}{m_C}\int \hat{\mathbf{A}}^2 d\mathbf{r}, \qquad (9)$$

$\mathbf{A}$ is the vector-potential of the electromagnetic field generated by the superconducting ring, $m_C$ and $2e$ is the mass and charge of the Cooper pair.

As discussed in section II, at low microwave intensities, satisfied (6), one can neglect the field correction to the ring energy. It means that the commutator of the potential energy of the Cooper pairs in the classical electromagnetic field $U_f$ and the Hamiltonian $H_0$ is small.

The evolution operator satisfies the equation:

$$i\hbar\frac{\partial}{\partial t}S(t,0) = \left[\exp\left(\frac{i}{\hbar}H_0 t\right)W\exp\left(-\frac{i}{\hbar}H_0 t\right) + U_f\right]S(t,0) \quad (10)$$

with $S(0,0) = 1$. In the first order over the operator $W$ we have:

$$S_1 = -\frac{i}{\hbar}\int_0^t dt_1 e^{\frac{i}{\hbar}H_0 t_1} W e^{-\frac{i}{\hbar}H_0 t_1} e^{-\frac{i}{\hbar}\int_0^t U_f dt_2}. \qquad (11)$$

It is easy to see that the action of $\exp\left(-\frac{i}{\hbar}\int_0^t U_f(t_1)dt_1\right)$ on the wave function (3) results in the change the phase of the single boson wave function which corresponds to (5).

### IV. TWO PHOTON EMISSION

Taking into account (9) and (11), the amplitude of the transition $m \to m_1$ is given by:

$$S_{if} = -\frac{i}{\hbar}\int_0^t W^{\mathbf{k}_1\mathbf{k}_2,0}_{m_1,m} e^{-\frac{it_1}{\hbar}(E_m - E_{m_1} - \hbar\omega_{k_1} - \hbar\omega_{k_2})} dt_1, \quad (12)$$

where

$$W^{\mathbf{k}_1\mathbf{k}_2,0}_{m_1,m} = \frac{2\hbar e^2}{mC\Omega_r} * \frac{(\mathbf{l}_{\mathbf{k}_1\sigma_1}\mathbf{l}_{\mathbf{k}_2\sigma_2})}{\varepsilon_0 \sqrt{\omega_{k_1}\omega_{k_2}}} \frac{\sin(\frac{d}{2}k_z)}{k_z k} * I_3(N), \quad (13)$$

$$k_z = k_1 \cos\theta_{k_1} + k_2 \cos\theta_{k_2},$$

$$I_3 = N\int_a^b \rho d\rho \int_0^{2\pi} d\varphi\, e^{iN[(m-m_1)\varphi - \xi_0 \rho \sin(\varphi)\sin(\omega_0 t)]} *$$
$$e^{-ik_1\rho\sin(\theta_{k_1})\cos(\varphi-\varphi_{k_1}) - ik_2\rho\sin(\theta_{k_2})\cos(\varphi-\varphi_{k_2})}. \quad (14)$$

Here $\mathbf{l}_{\mathbf{k}_1\sigma_1}$ and $\mathbf{l}_{\mathbf{k}_2\sigma_2}$ are the polarization vectors of the photons with the energies $\hbar\omega_{k_1}$ and $\hbar\omega_{k_2}$; $\theta_{k_1}, \theta_{k_2}$ and $\varphi_{k_1}, \varphi_{k_2}$ are the polar and azimuthal angles of their wave vectors.

Using (2) and (12)-(13), the probability of the transition channel $m \to m_1$ is written as:

$$w^t_{m_1 m} = \frac{2\pi}{\hbar} \sum_N \sum_{\mathbf{k}_1,\mathbf{k}_2} |c_{Nm}|^2 < |W^{\mathbf{k}_1\mathbf{k}_2,0}_{m_1 m}(N)|^2 >_{pol} *$$
$$\delta(E_m - E_{m_1} - \hbar\omega_{k_1} - \hbar\omega_{k_2}), \quad (15)$$

where $<...>_{pol} = (2\pi)^{-2} \sum_{\sigma_1,\sigma_2} \int_0^{2\pi} d\varphi_{l_1} \int_0^{2\pi} d\varphi_{l_2} ...$ means the average over the photon polarizations, $\varphi_{l_1}$ and $\varphi_{l_2}$ are the azimuthal angles of the polarization vectors of the photons.

For calculation of the integral in the right-hand side of (14) we used the procedure presented in [1]. Omitting these bulky calculations, we result in the expression for the probability of the transition (15) that is valid for the ring with the average number of the Cooper pairs $N > 10^9$ at their density $10^{21} cm^{-3}$ ( for mesoscopic rings it is only the first term of a series ):

$$w^t_{m_1 m} = w_0 \int_0^1 \gamma d\gamma \int_0^{\pi/2} \sin\theta_{k_1} d\theta_{k_1} \int_0^{\pi/2} d\varphi_{k_1} \int_{\pi/2}^{\pi} \frac{tg\theta^*_{k_2} d\varphi_{k_2}}{|\cos\varphi_{k_2}|}$$
$$\{G(\pi + \theta^*_{k_2})[D(\theta_{k_1},\theta^*_{k_2},\varphi_{k_1}-\varphi_{k_2}) + D(\theta_{k_1},\theta^*_{k_2},\varphi_{k_1}+\varphi_{k_2})] +$$
$$G(\theta^*_{k_2})[D(\pi - \theta_{k_1},\theta^*_{k_2},\varphi_{k_1}-\varphi_{k_2}) + D(\pi-\theta_{k_1},\theta^*_{k_2},\varphi_{k_1}+\varphi_{k_2})]\}$$
$$(16)$$

Here

$$w_0 = \frac{2^3 \alpha^2}{\pi^4} \left(\frac{2m_e}{mC}\right)^2 \frac{\lambdabar_e^2 c}{d^2(b-a)} \frac{3+(-1)^{m-m_1}}{\xi_0^2 R^2 - (m-m_1)^2},$$

$$G(\theta^*_{k_2}) = \left[\frac{\sin\left(\beta(\gamma(\cos\theta_{k_1}+\cos\theta^*_{k_2})-\cos\theta^*_{k_2})\right)}{\gamma(\cos\theta_{k_1}+\cos\theta^*_{k_2})-\cos\theta^*_{k_2}}\right]^2,$$

$\alpha$ is the fine structure constant, $\lambdabar_e$ is the Compton wavelength of the electron, $R(a < R < b)$ is the effective radius of the ring (it can be introduced because $(b-a)/b << 1$), $\gamma = \hbar\omega_{k_1}/E_0(m^2 - m_1^2)$ is the relative energy of the photon, $\beta = dE_0(m^2 - m_1^2)/2c\hbar$,

$$\theta^*_{k_2} = \arcsin\left(\frac{\gamma}{\gamma-1}\sin\theta_{k_1}\frac{\cos\varphi_{k_1}}{\cos\varphi_{k_2}}\right),$$

and $D$ is the correlation function,

$$D = 2|\cos\theta_{k_1}\cos\theta_{k_2}| + 4(1-|\cos\theta_{k_1}|)(1-|\cos\theta_{k_2}|)*$$
$$\left(1 + \frac{2\cos(\varphi_{k_1}-\varphi_{k_2})}{tg\theta_{k_1} tg\theta_{k_2}}\right) + 8\cos 2(\varphi_{k_1}-\varphi_{k_2})*$$
$$\frac{1-(1+\frac{1}{2}tg^2\theta_{k_1})|\cos\theta_{k_1}|}{tg^2\theta_{k_1}} * \frac{1-(1+\frac{1}{2}tg^2\theta_{k_2})|\cos\theta_{k_2}|}{tg^2\theta_{k_2}}.$$

The integrals in (16) are restricted by:

$$-1 \leq \frac{\gamma}{1-\gamma}\sin\theta_{k_1}\frac{\cos\varphi_{k_1}}{\cos\varphi_{k_2}} \leq 0. \quad (17)$$

It should be noted that the two-photon probability is proportional to the square of the fine structure constant. It is due to the "quantum" energy of the ring $E_0 \propto e^{-2}$.

Taking into account (5)-(6), in the expression (16) the low-frequency field strength is restricted by:

$$m - m_1 < \frac{2eE_0 R}{\hbar\omega_0} << m. \quad (18)$$

That is, the formula (16) is valid only for the decay channel $m \to m_1$ satisfying the conditions $\xi_0 a > (m-m_1)$ and $\xi_0 b > (m-m_1)$. These channels can be named as the allowed ones. According to (18), the higher the low-frequency field strength the greater number of the allowed transitions. In the case $\xi_0 b < (m-m_1)$ the decay probability for the channel $m - m_1$ is exponentially small, and this transition can be treated as forbidden.

V. NUMERICAL RESULTS AND DISCUSSIONS

Calculations of the two-photon transition probabilities were carried out for the rectangular cross-section rings with the outer radii $b = 5\mu$ and $16,7\mu$, the inner radius $a = 0.8b$, and the ring thickness was assumed $d = 800$ angstrom. The penetration depth is used the typical value for the type-II superconductors $\lambda = 2*10^3 A^0$. We used the results for the self-inductance $R_L = L/\mu_0$ of superconducting thin flat rings as a function of the quantity $a/b$, shown in Fig. 2 of [4]. Then, introducing the notation $R_L = \chi a$, we obtained $\chi(b=5\mu) = 7.1$ $E_0(b=5\mu) = 374$ meV and $\chi(b=16.7\mu) = 4.6$, $E_0(b=16.7\mu) = 173$ meV.

Let us estimate the intensity of the low-frequency microwave field with the frequency $\omega_0 = 10^{12} c^{-1}$ and $\xi_0 R = 3.5$. That is, the three transitions with $m - m_1 = 1,2,3$ are allowed. Using (6), we have $E_0 = 1.8 \hbar \omega_0 / eR$. For $R = 10\mu$ we obtained the intensity $\approx 1.3 mW/cm^2$.

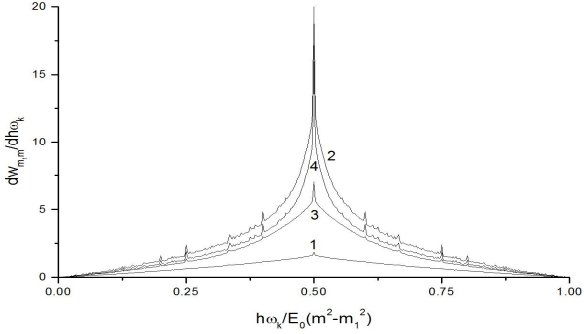

Fig.1. The energy distribution of photons emitted by the superconducting rings. Curve 1 - $b=5\mu$, $\xi_0 R = 1.3$, $m = 11$; curve 2 – $b = 5\mu$, $\xi_0 R = 4.3$, $m = 101$, $m - m_1 = 2$; curve 3 - $b = 16.7\mu$, $\xi_0 R = 1.3$, $m = 101$, $m - m_1 = 1$; curve 4 - $b = 16.7\mu$, $\xi_0 R = 5.3$, $m = 101$, $m - m_1 = 5$.

The energy spectra of the photons are demonstrated in Fig. 1. The distributions are always peaked up around the photon energy $\hbar \omega_p = E_0(m^2 - m_1^2)/2$. With the increasing of the fluxoid number in the initial state the peak width decreases sharply (curves 1 and 2 in Fig.1), that is, the probability that each of the two photons has the energy $\hbar \omega_p$, is increased. The reducing of the uncertainty in the energies of the photons can also be obtained by increasing the intensity of the microwave field that results in the increasing of the number of the allowed transitions (curves 3 and 4 in Fig 1.).

The polar angle distributions of the emitted photons are always peaked up around $\pi/2$, as shown in Fig. 2. The peak width decreases with increasing the change of the ring energy $E_0(m^2 - m_1^2)$. The increase of this energy can be obtained by decreasing the ring size, and increasing the fluxoid number in the initial state. Also, the narrowing of the peak can be achieved the increasing of the microwave field intensity (curves 1-4 in Fig. 2).

The azimuthal angle distributions of the emitted photons are always peaked up around angles $\varphi_{k_{1,2}} = \pi/2$, and $3\pi/2$ that corresponds to the $y$-axis along which the microwave field strength is directed. Note the spatial correlation in the emission of two photons, which is apparent from (17):

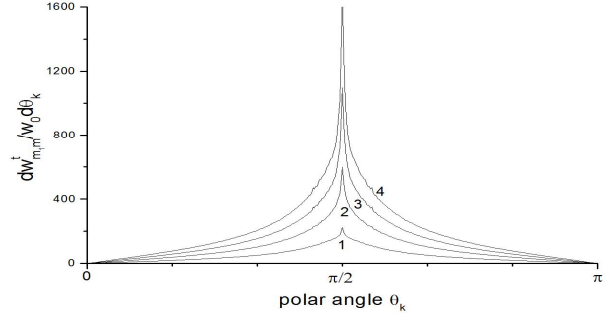

Fig. 2. The polar angle distributions of photons. Parameters: $b = 5\mu$, $\xi_0 R = 4.3$ and $m = 101$. Curve 1 – the decay channel $m - m_1 = 1$; curve 2 – $m - m_1 = 2$; curve 3 - $m - m_1 = 3$; curve 4 - $m - m_1 = 4$.

$\cos\varphi_{k_1} \cos\varphi_{k_2} \leq 0$. With account of the polar angle distributions discussed above, it means that the maximum probability of the two-photon emission meets the condition $\mathbf{k}_1 = -\mathbf{k}_2$. That is, two photons are emitted in nearly opposite directions, with the photon wave vector approximately perpendicular to the z-axis.

Table 1. The lifetimes (in seconds) for allowed two-photon transitions in the rings with $b = 5\mu$ and $m = 101$.

| $\xi_0 R$ | $\tau^t_{m-1,m}$ (75 eV) | $\tau^t_{m-2,m}$ (150 eV) | $\tau^t_{m-3,m}$ (223 eV) | $\tau^t_{m-4,m}$ (296 eV) | $\tau^t_{m-5,m}$ (s) (368 eV) |
|---|---|---|---|---|---|
| 1.3 | 0.078 | - | - | - | - |
| 2.3 | 0.482 | 0.035 | - | - | - |
| 3.3 | 1.111 | 0.186 | 0.067 | - | - |
| 4.3 | 1.965 | 0.392 | 0.338 | 0.033 | - |
| 5.3 | 3.106 | 0.671 | 0.705 | 0.168 | 0.069 |

The lifetimes $\tau^t_{m_1 m} = (w^t_{m_1 m})^{-1}$ for allowed transitions in the dependence on the amplitude of the microwave field given in units $\xi_0 R$, are presented in Table 1. Together with the lifetimes, the total energies of the emitted photons are shown. At low intensities the microwave field $\xi_0 R < 1$ allowed transitions are absent, and the superconducting current in the ring is persistent. For rings with $b = 16.7\mu$ the lifetimes are about one order of magnitude greater than the data presented.

[1] A.I. Agafonov, "Electromagnetic-field-induced decay of currents in thin-film superconducting rings with photons emission", arXiv:1107.2905, unpublished
[2] P.W. Anderson, "Considerations on Flow of Superfluid Helium", Rev. Mod. Phys., vol. 38, pp. 298-310, 1966.
[3] A. J. Leggett I and F. Sols, "On the Concept of Spontaneously Broken Gauge Symmetry in Condensed Matter Physics", Foundations of Physics, vol. 21, p. 353, 1991.
[4] E.H. Brandt and J.R. Clem, "Superconducting thin rings with finite penetration depth", Phys. Rev. B, vol. 69, p. 184509, 2004.